\documentclass[trackchanges,twocolumn]{aastex701}
\usepackage{booktabs}
\usepackage{xcolor}
\usepackage{amsmath}

\def\ra#1#2#3{#1$^{\rm h}$#2$^{\rm m}$#3$^{\rm s}$}
\def\dec#1#2#3{$#1^\circ#2'#3''$}

\newcommand{\twoten}{FRB\,20240210A}
\newcommand{\threetwelve}{FRB\,20240312D}
\newcommand{\twotwoseven}{FRB\,20250227A}
\newcommand{\onetwoseven}{FRB\,20211127I}
\newcommand{\threeonesix}{FRB\,20250316A}

\newcommand{\OII}{\ensuremath{[\rm{O}\,\textsc{ii}]}}
\newcommand{\OIII}{\ensuremath{[\rm{O}\,\textsc{iii}]}}
\newcommand{\NII}{\ensuremath{[\rm{N}\,\textsc{ii}]}}

\newcommand{\Halpha}{\ensuremath{{\rm H}\alpha}}
\newcommand{\Hbeta}{\ensuremath{{\rm H}\beta}}
\newcommand{\OH}{\ensuremath{12+\log(\mathrm{O/H})}}
\newcommand{\Hasb}{\ensuremath{\Sigma_{\mathrm{H}\alpha}}}

\begin{document}

\title{A spatially resolved view of Fast Radio Burst host metallicities with Integral-field Spectroscopy}

\author[0000-0003-3801-1496]{Sunil Simha}
  \email[show]{sunilsimha@uchicago.edu}
  \altaffiliation{NU-UC Brinson Fellow}
  \affiliation{Center for Interdisciplinary Exploration and Research in Astrophysics, Northwestern University, 1800 Sherman Avenue, Evanston, IL 60201, USA}
  \affiliation{Department of Astronomy and Astrophysics, University of Chicago, William Eckhardt Research Center, 5640 S Ellis Ave, Chicago, IL 60637, USA}

\author[0000-0001-8813-4182]{Hsiao-Wen Chen}
    \email{hwchen@uchicago.edu}
    \affiliation{Department of Astronomy and Astrophysics, University of Chicago, William Eckhardt Research Center, 5640 S Ellis Ave, Chicago, IL 60637, USA}

\author[0000-0002-7374-935X]{Wen-fai Fong}
    \email{wfong@northwestern.edu}
    \affiliation{Center for Interdisciplinary Exploration and Research in Astrophysics, Northwestern University, 1800 Sherman Avenue, Evanston, IL 60201, USA}

\author[0000-0002-7332-2751]{Sarah Hughes}
    \email{slhughes@mit.edu}
    \affiliation{MIT Kavli Institute for Astrophysics and Space Research, Massachusetts Institute of Technology, 77 Massachusetts Ave, Cambridge, MA 02139, USA}

\author[0000-0002-9363-8606]{Yuxin Dong}
    \email{yuxin.dong@northwestern.edu}
    \affiliation{Center for Interdisciplinary Exploration and Research in Astrophysics, Northwestern University, 1800 Sherman Avenue, Evanston, IL 60201, USA}

\author[0000-0002-7738-6875]{J. Xavier Prochaska}
    \email{xavier@ucolick.org}
    \affiliation{Department of Astronomy and Astrophysics, University of California, Santa Cruz, 1156 High Street, Santa Cruz, CA 95060, USA}
    \affiliation{Kavli Institute for the Physics and Mathematics of the Universe (Kavli IPMU), 5-1-5 Kashiwanoha, Kashiwa, 277-8583, Japan}
    \affiliation{Division of Science, National Astronomical Observatory of Japan, 2-21-1 Osawa, Mitaka, Tokyo 181-8588, Japan}

\author[0000-0002-1883-4252]{Nicolas Tejos}
    \email{nicolas.tejos@pucv.cl}
    \affiliation{Instituto de F\'isica, Pontificia Universidad Cat\'olica de Valpara\'iso, Casilla 4059, Valpara\'iso, Chile}

\author[0000-0003-0307-9984]{Tarraneh Eftekhari}
    \email{teftekhari@northwestern.edu}
    \affiliation{Center for Interdisciplinary Exploration and Research in Astrophysics, Northwestern University, 1800 Sherman Avenue, Evanston, IL 60201, USA}

\author[0000-0002-5025-4645]{Alexa C. Gordon}
    \email{agordon@caltech.edu}
    \altaffiliation{Brinson Prize Fellow}
    \affiliation{Division of Physics, Mathematics, and Astronomy, California Institute of Technology, Pasadena, CA 91125, USA}

\author[0000-0003-4501-8100]{Stuart D. Ryder}
    \email{Stuart.Ryder@mq.edu.au}
    \affiliation{School of Mathematical and Physical Sciences, Macquarie University, Sydney, NSW 2109, Australia}
    \affiliation{Astrophysics and Space Technologies Research Centre, Macquarie University, Sydney, NSW 2109, Australia}

\author[0000-0002-4623-5329]{Mawson W. Sammons}
    \email{mawson.sammons@lam.fr}
    \affiliation{Laboratoire d'Astrophysique de Marseille, Aix-Marseille Univ., CNRS, CNES, Marseille, France}

\author[0000-0002-8376-1563]{Alice P. Curtin}
    \email{alice.curtin@mail.mcgill.ca}
    \affiliation{Department of Physics, McGill University, 3600 rue University, Montr\'eal, QC H3A 2T8, Canada}
    \affiliation{Trottier Space Institute, McGill University, 3550 rue University, Montr\'eal, QC H3A 2A7, Canada}
    \affiliation{Anton Pannekoek Institute for Astronomy, University of Amsterdam, Science Park 904, 1098 XH Amsterdam, The Netherlands}

\author[0000-0001-9345-0307]{Victoria M.~Kaspi}
  \email{victoria.kaspi@mcgill.ca}
  \affiliation{Department of Physics, McGill University, 3600 rue University, Montr\'eal, QC H3A 2T8, Canada}
  \affiliation{Trottier Space Institute, McGill University, 3550 rue University, Montr\'eal, QC H3A 2A7, Canada}

\author[0000-0003-2116-3573]{Adam E. Lanman}
    \email{alanman@mit.edu}
    \affiliation{MIT Kavli Institute for Astrophysics and Space Research, Massachusetts Institute of Technology, 77 Massachusetts Ave, Cambridge, MA 02139, USA}
    \affiliation{Department of Physics, Massachusetts Institute of Technology, 77 Massachusetts Ave, Cambridge, MA 02139, USA}

\author[0000-0002-4209-7408]{Calvin Leung}
\affiliation{Miller Institute for Basic Research, University of California, Berkeley, CA 94720, United States}
\affiliation{Department of Astronomy, University of California, Berkeley, CA 94720, United States}
\email{calvin_leung@berkeley.edu}

\author[0000-0002-4279-6946]{Kiyoshi W. Masui}
    \email{kmasui@mit.edu}
    \affiliation{MIT Kavli Institute for Astrophysics and Space Research, Massachusetts Institute of Technology, 77 Massachusetts Ave, Cambridge, MA 02139, USA}
    \affiliation{Department of Physics, Massachusetts Institute of Technology, 77 Massachusetts Ave, Cambridge, MA 02139, USA}

\author[0000-0002-8912-0732]{Aaron B. Pearlman}
    \email{aaron.b.pearlman@mit.edu}
    \altaffiliation{NASA Hubble Fellow}
    \affiliation{MIT Kavli Institute for Astrophysics and Space Research, Massachusetts Institute of Technology, 77 Massachusetts Ave, Cambridge, MA 02139, USA}
    \affiliation{Department of Physics, McGill University, 3600 rue University, Montr\'eal, QC H3A 2T8, Canada}
    \affiliation{Trottier Space Institute, McGill University, 3550 rue University, Montr\'eal, QC H3A 2A7, Canada}

\author[0000-0002-7285-6348]{Ryan~M.~Shannon}
    \email{ryanmshannon@gmail.com}
    \affiliation{Centre for Astrophysics and Supercomputing, Swinburne University of Technology, Hawthorn, VIC 3122, Australia}

\begin{abstract}

Magnetar sources are the leading theory for Fast Radio Burst (FRB) progenitors. If magnetars are indeed the dominant sources, we may thus expect any metallicity preference for magnetar formation to be reflected in the FRB site in host galaxies. Here, we leverage Integral Field Unit (IFU) spectroscopy to spatially-resolve gas-phase metallicity distribution in five FRB host galaxies, \twoten, \threetwelve, \twotwoseven, \onetwoseven, and \threeonesix, at $z<0.05$, including the FRB site, with the O3N2 index. 
We observed the first four sources with the newly commissioned Large Lenslet Array Magellan Spectrograph (LLAMAS) IFU on the Magellan Telescopes and used archival Keck/KCWI data for \threeonesix. 
We find ISM metallicity variations of at least 0.3 dex within individual hosts for the majority of the sample, with median values spanning $12+\log({\rm O}/{\rm H})\approx8.5$--8.7. At the burst sites, four have metallicities of $12+\log({\rm O}/{\rm H})\approx8.5$--8.9, while the fifth has a metallicity below 8.5. These findings reveal substantial metallicity variations both among FRB host galaxies and within the local environments of the bursts. Consequently, our study shows that longslit-based global estimates of host galaxy metallicity do not always represent the true burst-site distributions. We therefore advocate for IFU follow-up of larger host samples to ultimately provide FRB-site relevant observational constraints on metallicity preferences for FRB occurrence.   

\end{abstract}



\section{Introduction}

Fast radio bursts (FRBs) are millisecond-duration extragalactic radio transients whose progenitors remain observationally elusive, and the properties of their host galaxies have become one of the primary tools for constraining candidate formation channels. Systematic studies of FRB hosts have largely focused on global host properties estimated from integrated photometry or spectroscopy. These have inferred that the population is diverse: FRBs mostly occur in actively star-forming galaxies, broadly consistent with a picture in which most progenitors form promptly after core collapse of massive stars \citep[e.g.][]{Heintz+2020, Bhandari+2022, Gordon+2023, Sharma+2024}. \citet{Sharma+2024} additional reported a preference for massive hosts or a deficit of low-mass hosts ($\log(M_*/M_\odot)\lesssim9$; total sample size $N_{\text{FRB}}=20$) at $z\leq0.2$ relative to the cosmic star-formation rate. Adopting a population-level mass--metallicity relation, they translated this deficit into an inferred suppression of FRB occurrence at $\OH<8.08^{+0.61}_{-0.47}$. Subsequently, \citet{Yamasaki+2026} directly estimated oxygen abundances with the O3N2 index \citep{Pettini+2004} for a larger volume-limited sample of FRB hosts ($N_{FRB}=27$; $0.01\leq z\leq0.48$) with emission lines but found no evidence for a metallicity lower bound. Additionally, \citet{Yamasaki+2026, Moroianu+2026} and \citet{Muller+2026} find hosts with low metallicity ($\OH\sim 8.0$). Most recently, \citet{Dong+2026} examine a larger sample of FRB hosts including five new dwarf hosts and the host sample of \citet{Sharma+2024} and find evidence for metallicity preferences for FRBs. Specifically, they report that hosts of non-repeating FRBs occur preferentially at $\OH\gtrsim8.35$ while repeaters prefer $\OH\lesssim8.35$. Additionally, FRBs associated with persistent radio sources (PRSs) prefer dwarfs with $\OH\lesssim8.15$.

On the theory front, a magnetar origin is the leading source model for FRB progenitors \citep{Bochenek+2020c, CHIME+2020, ZhangB2023}, and any metallicity preference for magnetar formation should be reflected in FRB occurrence in varying galaxy types. \citet{Sharma+2024} have theorized that isolated massive stars can preferentially evolve into neutron stars, as opposed to black holes, in a solar or higher metallicity environment due to more efficient stellar winds \citep{Heger+2002}. They also speculate that in binary systems, higher metallicities can more reliably trigger faster Roche-lobe overflow \citep{Klencki+2020}, which can lead to massive-star mergers that could lead to more frequent magnetars \citep{Frost+2024}. In contrast, \citet{Dong+2026} propose an alternative mechanism, specifically for repeating FRBs, wherein low-metallicity environments can produce faster-rotating stars and subsequently faster magnetars from lower mass loss in low-metallicity environments \citep{WoosleyBloom2006}. Supporting evidence comes from \citet{Schootemeijer+2022}, who find a high fraction of extreme OBe stars in metal-poor dwarfs. Thus, the exact mechanism that drives metallicity preference is debated.

\begin{deluxetable*}{lccccccccc}[t!]
\tabletypesize{\footnotesize}
\tablecaption{Sample properties \label{tab:sample}}
\tablecolumns{10}
\tablewidth{0pt}
\tablehead{
\colhead{} &
\colhead{} &
\colhead{R.A. (host)} &
\colhead{Dec. (host)} &
\colhead{$\theta_{\rm gal}$} &
\colhead{R.A. (FRB)} &
\colhead{Dec. (FRB)} &
\colhead{$a_{\rm err}$} &
\colhead{$\delta$R} &
\colhead{Stellar Mass} \\
\colhead{FRB} &
\colhead{$z$} &
\colhead{(J2000)} &
\colhead{(J2000)} &
\colhead{($\arcsec$)} &
\colhead{(J2000)} &
\colhead{(J2000)} &
\colhead{($\arcsec$)} &
\colhead{(kpc)} &
\colhead{$\log(M_\star/M_\odot)$}
}
\startdata
20240210A & 0.0239 & \ra{00}{35}{06.51} & $-$\dec{28}{16}{19.00} & 16.3 & \ra{00}{35}{07.10} & $-$\dec{28}{16}{14.03} & 0.33 & 5 & $10.3$ \\
20240312D & 0.0495 & \ra{03}{26}{29.50} & $-$\dec{54}{35}{57.20} & 5.9 & \ra{03}{26}{29.61} & $-$\dec{54}{36}{03.54} & 1.30 & 6 & $10.7$ \\
20250227A & 0.0314 & \ra{07}{33}{30.17} & \dec{$+$15}{13}{34.70} & 12.4 & \ra{07}{33}{30.66} & \dec{$+$15}{13}{23.77} & 0.45 & 8 & $8.9$ \\
20211127I & 0.0469 & \ra{13}{19}{13.97} & $-$\dec{18}{50}{16.07} & 17.9 & \ra{13}{19}{14.10} & $-$\dec{18}{50}{17.00} & 0.51 & 2 & $9.5$ \\
20250316A & 0.0069 & \ra{12}{09}{47.31} & \dec{$+$58}{50}{57.25} & 23.8 & \ra{12}{09}{44.31} & \dec{$+$58}{50}{56.71} & 0.09 & 5$^{a}$ & $9.2$ \\
\enddata
\tablecomments{FRB localizations and redshifts for CRAFT detected FRBs are from \citealt{Shannon2025}. \threeonesix\ was reported by \citet{RBFLOAT_2025}. \twotwoseven\ is reported by \citet{Dong+2026}. $\theta_{\rm gal}$ is the host galaxy's angular size (Pan-STARRS Kron radius, or the DECaLS half-light radius for FRB\,20240312D). $a_{\rm err}$ is the localization uncertainty semi-major axis. Galactocentric offsets $\delta$R are the host-to-FRB angular separation converted to a projected physical offset using the Planck18 angular diameter distance at the host redshift. $^{a}$For FRB20250316A (host NGC4141, $z=0.00689$), the redshift-derived distance is unreliable due to peculiar velocity; we instead adopt $D=40.8$ Mpc, the mean of two redshift-independent Tully--Fisher distances \citep{2014MNRAS.444..527S, 2016AJ....152...50T} from the NED Distance database. Stellar masses are compiled from the literature: \twoten\ and \threetwelve\ are from \citealt{Gordon2025}; \onetwoseven\ is from \citealt{Gordon+2023}, as tabulated by \citealt{Gordon2025}; \twotwoseven\ is from Prospector SED fits presented by \citet{Dong+2026}; and \threeonesix\ is from the NED Local-Volume sample \citep{Cook+2023_NED}, as reported by \citealt{RBFLOAT_2025}.}
\end{deluxetable*}

\begin{figure*}
    \centering
    \includegraphics[width=0.8\textwidth]{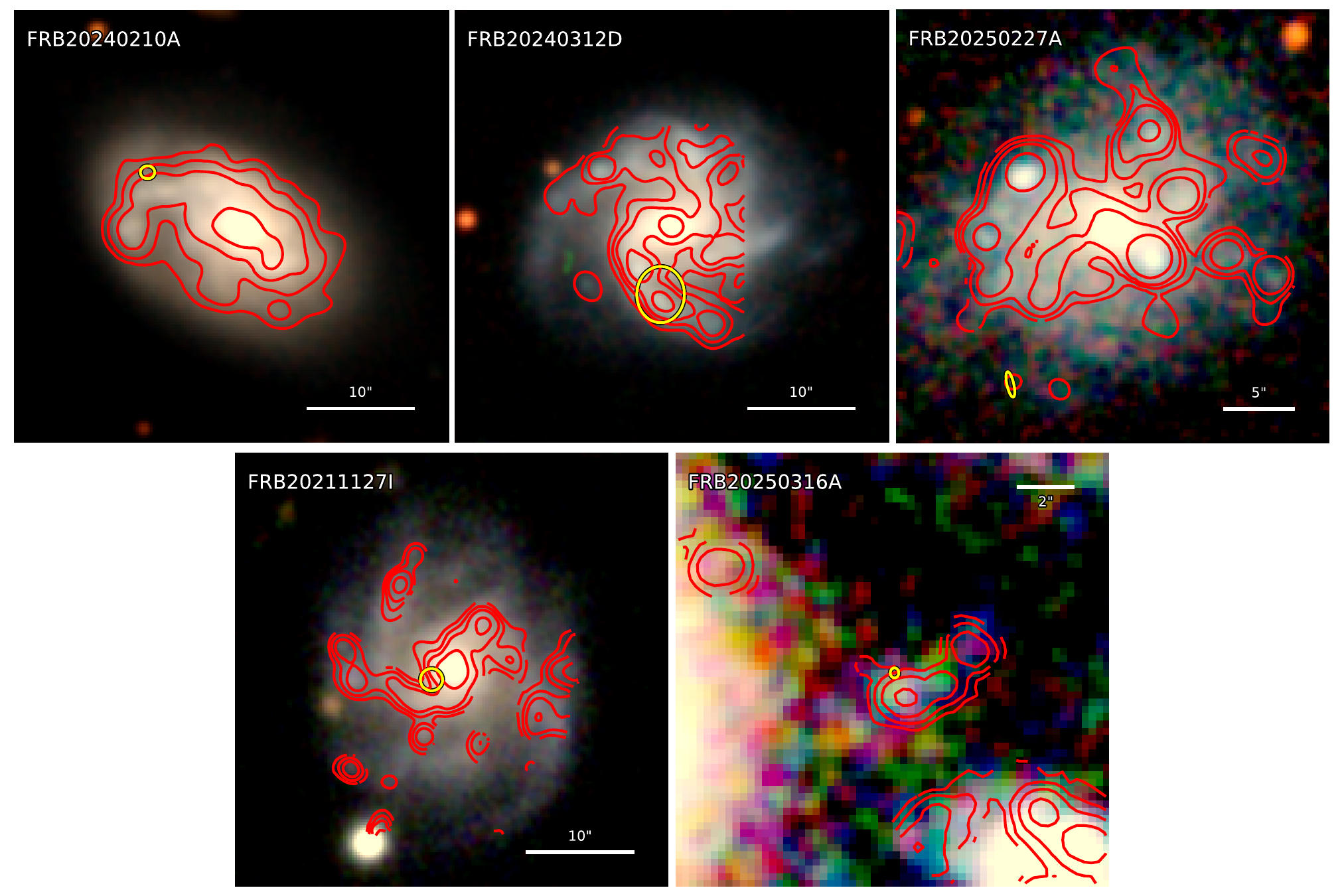}
    \caption{Optical false-color images showcasing the host galaxies of FRBs in our sample: \twoten, \threetwelve, \twotwoseven, \onetwoseven\ and \threeonesix. The color images are produced from $g$, $r$, $z$ cutouts from the Legacy Survey archive \citep[][]{DECALs_2019} for the first three hosts and Pan-STARRS $g$, $r$, $i$ for the other two due to the sky coverage of the respective surveys. The yellow ellipses mark the 2$\sigma$ FRB localization regions \citep{Shannon+2025, RBFLOAT_2025}. Red contours trace the \Halpha\ surface brightness measured from our LLAMAS and KCWI IFU line maps, drawn at four levels spanning the 75th--99th percentile of the positive \Halpha\ flux in each map. We note in the case of \threetwelve, our data did not cover the entire galaxy due to a pointing error; hence the sharp edges to the west and the north of the image. In the case of \threeonesix\, the data covers a $8''\times20''$ footprint of the host, NGC 4141, and thus we only show the region around the FRB location.}
    \label{fig:sample}
\end{figure*}

Regardless of the exact mechanism, granting that a metallicity-sensitive channel exists for magnetar formation, an FRB is a local event which is only sensitive to its immediate neighborhood, and not necessarily the host galaxy's global properties, i.e., the host metallicity inferred from integrated light might not be representative of the local value at the progenitor formation site. Metallicity gradients spanning $\sim0.3$ dex are well-established in stacks of Sloan Digital Sky Survey (SDSS) galaxies, with individual galaxies exhibiting even broader spatial variations \citep[e.g.][]{Zaritsky+1994, Ryder+1995, Belfiore+2017, Lian+2018}. More directly, \citet{Pessi+2023} find that galaxies hosting core-collapse supernovae (CCSNe), a magnetar-formation channel, can show different metallicity values at the transient site than the global host value. Thus, there is a clear need to characterize FRB hosts at the burst site, ideally with integral field unit (IFU) observations. 

Though the FRB literature does not lack IFU observations nor inferences of host properties from them, \citep[e.g.,][]{Simha+2020, Chittidi+2021, Khrykin+2024, Bernales-Cortes+2025, Jahns-Schindler+2026}, only two studies by \citet{Tendulkar+2021} and \citet{RBFLOAT_2025} characterize local metallicity for FRB20180916B and \threeonesix\ respectively. Both studies are comparable with respect to FRB localization ($\sim50$ mas) and offset from the nearest star-forming region ($\sim200$ pc) to the FRB. They find that the local gas-phase metallicities are 8.4 and 8.65, respectively. Interestingly, these hosts also have global metallicity estimates from longslit observations. \citet{Marcote+2020} and \citet{Gordon+2023} estimate $\OH=8.8$--8.9 for FRB20180916B using Gemini and Keck longslit observations, 0.4--0.5 dex different from the local value. \citet{Dong+2026} estimate 8.64 for \threeonesix. While it seems apparently closer to the IFU estimate, \citet{RBFLOAT_2025} use the R23 diagnostic as opposed to O3N2 from \citet{Dong+2026}. R23 can systematically produce $0.1$--0.5 dex higher values \citep[e.g., ][]{Wisz+2025}; thus, there is likely a difference between global and site values for \threeonesix as well. We discuss \threeonesix\ further below, as it is re-analyzed in our study, but even with two hosts, it is apparent that the FRB-site metallicity is not always well-represented by global values.

Over the last five years, various FRB experiments \citep[]{Sharma+2024, Shannon+2025, Dong+2026} have built a small ($\lesssim10$) sample with good localization precision ($\lesssim1$ arcsec) in nearby hosts ($z\lesssim0.05$). Thus, for the first time, we are able to analyze the FRB-site metallicity as a sample. In this work, we present a study featuring data from the newly commissioned Large Lenslet Array Magellan Spectrograph (LLAMAS; \citealt{Furesz+2020}) mounted on the Magellan Baade Telescope to measure spatially resolved gas-phase metallicity, via the O3N2 index, at the precise position of four well-localized FRBs within their host galaxies.  This paper is organized as follows: Section \S\ref{sec:sample} describes our sample selection. Section \S\ref{sec:analysis} describes our emission-line fitting and metallicity-mapping procedure; \S\ref{sec:results} presents the resulting local metallicity maps and \S\ref{sec:discussion} discusses the implications of our results and future outlook. Throughout, we adopt a flat $\Lambda$CDM cosmology with Planck18 parameters \citep{Planck18}.

\section{Sample definition and IFU Observations}
\label{sec:sample}
\begin{deluxetable*}{lcccccc}[t!]
\tablecaption{Observation log \label{tab:obs}}
\tablecolumns{7}
\tablewidth{0pt}
\tablehead{
\colhead{FRB} &
\colhead{UT Date} &
\colhead{$N_{\rm exp}$} &
\colhead{$t_{\rm tot}$} &
\colhead{$\Sigma_{10\sigma}(\Halpha)$} &
\colhead{$\theta_{\rm PSF}$} &
\colhead{$\theta_{\rm PSF}$} \\
\colhead{} &
\colhead{} &
\colhead{} &
\colhead{(s)} &
\colhead{($10^{-17}\,{\rm erg\,s^{-1}\,cm^{-2}\,arcsec^{-2}}$)} &
\colhead{(\arcsec)} &
\colhead{(kpc)}
}
\startdata
20240210A & 2025 Dec 17 & 4 &  3600 & 0.43 & 1.8 & 0.90 \\
20240312D & 2025 Dec 17 & 4 &  3600 & 0.71 & 1.6 & 1.6 \\
20250227A & 2025 Dec 17 & 6 &  5400 & 0.37 & 1.8 & 1.2 \\
20211127I & 2026 Jul 4  & 2 &  1800  & 1.1 & 2.2 & 2.1 \\
20250316A* & 2025 Mar 23 & 2 (6) & 2060 (1800) & 5.1 & 0.7 & 0.13 \\
\enddata
\tablecomments{$N_{\rm exp}$ is the number of dithered science exposures combined into the final cube for each field, and $t_{\rm tot}=N_{\rm exp}\times 900s$ is the total on-source integration time. \onetwoseven\ observations were shortened by poor weather, and only two exposures were obtained. $\Sigma_{10\sigma}(\Halpha)$ is the faintest \Halpha\ surface brightness surviving each field's own per-spaxel S/N$\,\geq\,$10 detection threshold -- the same threshold used to build the \Halpha\ line maps shown in the figures. The limits do not scale well with exposure time because the per-spaxel formal error depends on reduction artifacts at the observed wavelength of \Halpha. Additionally, the KCWI cube for \threeonesix\ has smaller spaxel size compared to our LLAMAS cubes. We describe our masking procedure to filter out these artifacts in \S\ref{subsec:linefit}. $\theta_{\rm PSF}$ is the point-spread-function FWHM at 6200\,\AA, measured from the same-night standard star closest in airmass. We also convert this to a projected physical distances at the host distance to show our resolving power. FRB\,20250316A* has Keck KCWI/KCRM data from\citet{RBFLOAT_2025}; $N_{\rm exp}$ and $t_{\rm tot}$ give the KCWI (blue) value with the KCRM (red) value in parentheses.}
\end{deluxetable*}

Table \ref{tab:sample} lists the sources in our sample consisting of five apparently non-repeating FRBs with $\lesssim1''$  localizations. Three of our targets, \twoten, \threetwelve, and \onetwoseven\ were discovered and localized by the ASKAP/CRAFT experiment \citep{Shannon+2025}, while \twotwoseven\ and \threeonesix\ were detected and localized by the CHIME/FRB collaboration with the full Outrigger Array \citep[][]{outriggers_2025, Dong+2026}. These have well-associated nearby host galaxies: i.e., having a posterior value of $>0.99$ from the Probabilistic Association of Transients to their Hosts \citep[PATH ][]{PATH_2021} framework and host redshift $z<0.05$. The angular sizes of all the hosts are $\gtrsim5\times$ the $1\sigma$ FRB localization uncertainties. At the host distance, the $1\sigma$ FRB localization uncertainty corresponds to $\lesssim 1$ kpc in projected offset distance, thus ensuring that we are characterizing the environment local to the FRB. For this study, we observed four FRB hosts listed in Table~\ref{tab:sample} with the Magellan Baade/LLAMAS IFU \citep[P.I.: Simcoe;][]{Furesz+2020, Frostig+2020}. Table~\ref{tab:obs} summarizes the total integration times we obtained for each field. We dithered to cover gaps left by the two damaged blue detectors (detectors 1A and 4A). LLAMAS is a wide-field ($37''\times 37''$) fiber-fed IFU with a spectral resolution of $R\sim 2000$ and wavelength coverage of 3500--9700 \AA. The IFU has three channels: blue, green and red, with separate detectors to maximize photon counts. Each channel has eight detectors, among which the spectra from 2400 fibers are shared. Each detector collects light from a spatially connected section. The fibers have a pitch of $0.75''$.

For \threeonesix\ we re-analyze the Keck/KCWI data from \citet{RBFLOAT_2025} with the same methodology as the other fields. We refer the reader to the aforementioned work for details regarding the observations and reduction process of the KCWI data. However, we note here that the available data only covers a footprint of $8''\times20''$ of the host, NGC~4141. This is significantly smaller than the full size of the host, as its half-light radius is $\sim24''$. We explicitly mention how this coverage affects our analysis and interpretation where relevant below.

Our host stellar masses range from $10^{8.9}$--$10^{10.7}~M_\odot$ as compiled from the literature and listed in Table \ref{tab:sample}.

\section{Methods}
\label{sec:analysis}

\begin{figure*}
\centering
\includegraphics[width=\textwidth]{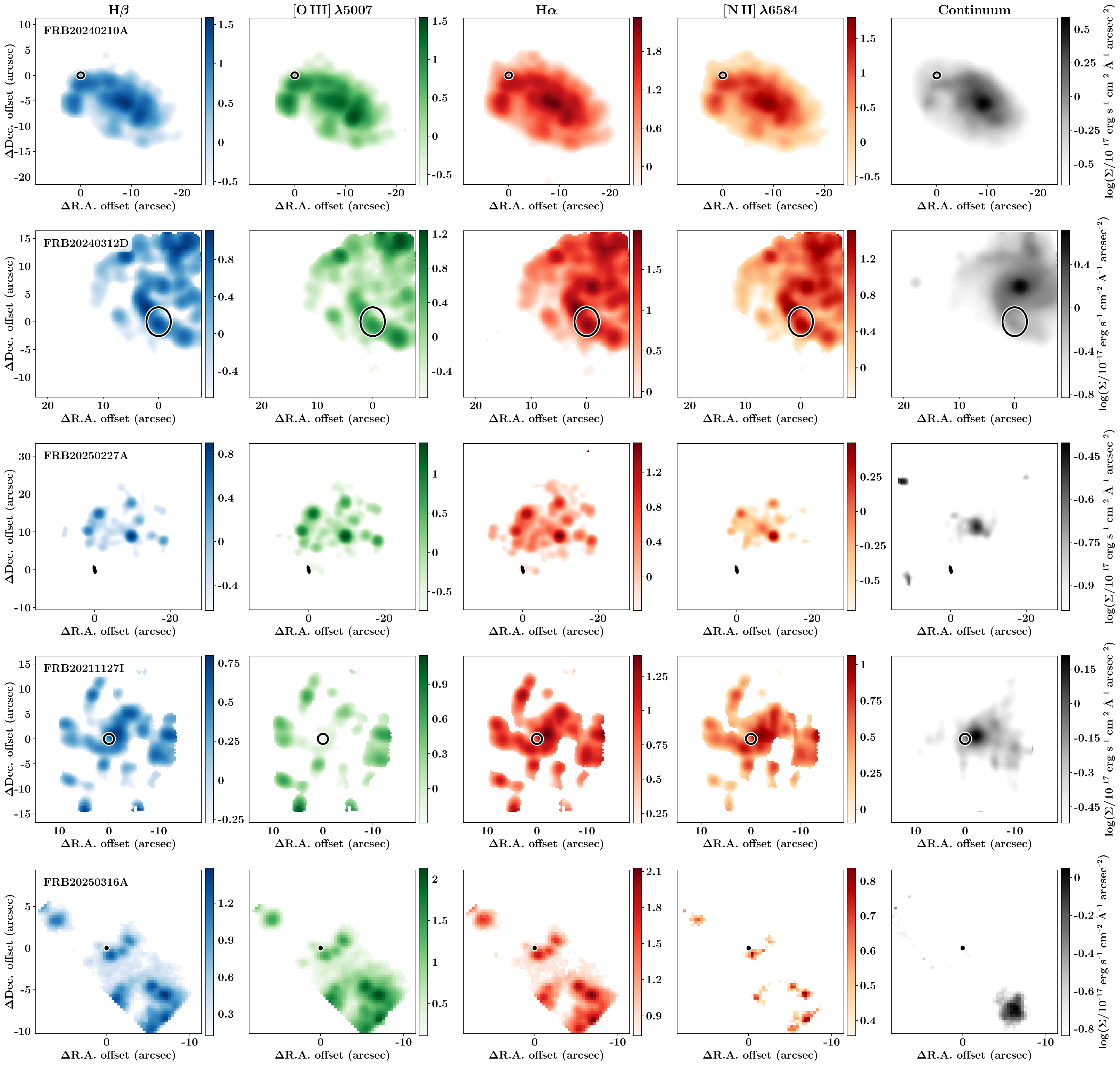}
\caption{
Surface brightness maps of \Hbeta, \OIII\,$\lambda5007$, \Halpha, \NII\,$\lambda6584$, and the stellar continuum for all five  FRB host galaxies
(one row per target). The color scale is logarithmic for all panels. The coordinate axis is centered on the FRB and shows relative offsets in arcseconds. Each line emission map is first restricted to the host galaxy's
footprint, defined by $\mathrm{S/N} \geq 10$ in \Halpha\ (for \onetwoseven, additionally requiring $\mathrm{S/N} \geq 5$ in \Hbeta\ to suppress a banding artifact; \S\ref{subsec:linefit}), and
within that footprint, a spaxel is shown only if the plotted line itself has $\mathrm{S/N} \geq 3$. The continuum map corresponds to emission at the $\geq2\sigma$ level. The black ellipse in each panel marks the FRB's $2\sigma$ positional uncertainty region. As mentioned before, the data for \threetwelve\ covers only the portion of the galaxy shown here. For \threeonesix\, we show only a $5''\times5''$ region around the FRB centroid to highlight the $\sim1$ kpc environment around the extremely precisely localized FRB.}
\label{fig:narrowband}
\end{figure*}

\subsection{Data reduction and cube construction}
The data were reduced using our GitHub fork of the LLAMAS-Pyjamas pipeline\footnote{https://github.com/SunilSimha/llamas-pyjamas} (Hughes et al. 2026; in prep). The fork incorporates features from the \texttt{rs-dev} branch\footnote{commit:c09e318525ec0590ceda44c50704ea4e379d0a2d of the upstream repository: https://github.com/mit-kavli-institute/llamas-pyjamas} of the parent repository but adds code for post-processing. The base pipeline performs bias and dark subtraction, flat fielding, wavelength calibration, sky subtraction, and cube construction for individual exposures.  A detailed description of the post-processing steps is provided in Appendix \ref{appendix:postprocessing}. Briefly, these include flux calibration, co-addition of cubes, and astrometric correction. We adapted code from the PypeIt package \citep{pypeit} to flux-calibrate the cubes using the standard star observations taken on the nights when the sources were observed. Then, for each FRB host, we reproject and co-add cubes from the individual exposures after registering them to a common WCS. Finally, we extract a narrow-band image (10 \AA\ bandwidth) centered on the galaxy \Halpha\ emission and align this to archival imaging from the Legacy Survey DR10 \citep{DECALs_2019} or Pan-STARRS DR2 \citep{Flewelling+2020}, which, in turn, are calibrated to \citet{GaiaDR2}. The results are flux-calibrated, coadded, and astrometrically calibrated cubes. Figure~\ref{fig:sample} illustrates the \Halpha\ emission from our final cubes as contours overlaid on false-color images of the hosts.

For the remainder of the analysis, we use the final cubes from the green channel alone, which includes the \Hbeta, \OIII$\lambda\lambda4959,5007$, \Halpha, and \NII$\lambda\lambda6548,6583$ emission lines. We note that we detect \OII$\lambda3726,3729$ emission in the blue channel for all sources but refrain from using it in our analysis below, as we observe significant artifacts from coadding the blue cubes with gaps from the two missing detectors.

Table \ref{tab:obs} lists the \Halpha\ surface brightness limit of each field's combined cube, defined as the background median plus 10 times the background standard deviation (details in the next subsection). The limits do not scale well with exposure time because the line flux estimation is affected by reduction artifacts which produce a non-Gaussian background spaxel distribution in a narrow-band image. This is particularly true for \onetwoseven\ where, since we only had two exposures which were not spatially dithered, vertical banding artifacts were above the 5$\sigma$ level. We therefore apply a thorough masking procedure (described further below) to filter out these artifacts and isolate the galaxy light. 

\subsection{Emission-line maps}
\label{subsec:linefit}
We obtain line emission maps for \Hbeta, \OIII\,$\lambda\lambda4959,5007$, \Halpha, and \NII\,$\lambda\lambda6548,6584$. We fit each spaxel of every coadded science cube's spectrum independently, producing the spatially resolved (effective PSF FWHM $\lesssim 2''$) flux maps shown in Figure \ref{fig:narrowband}. At each spaxel, two wavelength slabs corresponding to emission line groups -- \Hbeta+\OIII\ (restframe wavelength range 4760-5110 \AA) and the \Halpha+\NII (restframe wavelength range 6490-6620 \AA)  -- were fit independently with a cubic polynomial continuum plus a sum of Gaussian line profiles, with all lines in a given complex sharing one common redshift and line width as they are spectrally unresolved. For \Hbeta, we also fit a Gaussian with a negative amplitude to model the Balmer absorption from the stellar components of the galaxy spectrum. We tied the flux ratio of line doublets, whose relative strength is fixed by atomic transition probabilities (the two \OIII\ and two \NII\ lines), to their theoretical flux ratio rather than fit them independently. We propagated formal flux uncertainties for every line directly from each fit's covariance matrix.

To reject spurious emission from reduction or fitting artifacts, we restrict fits to the galaxy footprint, defined by \Halpha\ at $\mathrm{S/N}\geq10$ (i.e., spaxels not clearing this threshold are not considered for any other emission line fit). For \onetwoseven\ only, we additionally require \Hbeta\ at $\mathrm{S/N}\geq5$ -- a second, independent line whose own footprint suppresses a coherent banding artifact otherwise present in that target's green-channel cube. Within that footprint, a spaxel's flux for any individual line is retained only where that line itself independently clears $\mathrm{S/N}\geq3$. We show the emission line and continuum maps (from the wavelength slab containing \Halpha) in surface-brightness units in Figure \ref{fig:narrowband}.

\subsection{Metallicity maps}
Our primary goal is to obtain resolved measurements of a host galaxy's gas-phase metallicity at the FRB position. To this end, we employ the O3N2 index \citep{Pettini+2004} with the calibration from \citet{Nakajima+2022} with coefficients from their Table 2 for the \OH\ range of 7.59--8.9. i.e.
\begin{equation}
    \begin{aligned}
        \rm O3N2 =~&0.226 \\
        - &4.71(\OH -8.69)\\
        - &2.138(\OH -8.69)^2\\
    \end{aligned}
\end{equation}
where O3N2 = $\log((\OIII5007/\Hbeta) / (\NII6584/\Halpha))$.

We use our line maps to generate the gas-phase O3N2 maps as shown in panels 1-5 of Figure \ref{fig:metallicity_maps}. The O3N2 index is our preferred diagnostic as all the strong lines from \Hbeta\ to \NII\ are well-detected for all our sources. Since the diagnostic relies on the ratios \OIII/\Hbeta\ and \NII/\Halpha\, we avoid any potential dust reddening and flux calibration or flat-fielding issues between the two wavelength slabs analyzed that are separated by $\sim1500$ \AA.

\subsection{FRB-site estimates}
From our 2D maps, we estimate (1) the \Halpha\ surface brightness (\Hasb) and (2) the metallicity at the FRB location informed by the PSF size. We weight spaxels by a Gaussian whose FWHM corresponds to the PSF size listed in Table \ref{tab:obs} to estimate the mean value and the scatter among spaxels for the uncertainty. For \OH\, \NII\ is the limiting factor as it is the faintest line across all galaxies in our sample. For \twotwoseven, where no spaxel inside the FRB localization has an \NII\ detection, we instead report an upper limit based on the $3\sigma$ limit on \NII\,$\lambda6584$.

\section{Results}
\label{sec:results}
%
%

\begin{table*}
\centering
\footnotesize
\caption{O3N2 gas-phase metallicity statistics for the five FRB host galaxies.}
\label{tab:metallicity_stats}
\begin{tabular}{l|cc|ccc}
\hline\hline
 & \multicolumn{2}{c|}{Host} & \multicolumn{3}{c}{FRB location} \\
FRB & Median O3N2 & 95\% O3N2 interval & Mean O3N2 & Std. dev. O3N2 & $N\sigma$ from \\
    & ($12+\log_{10}(\mathrm{O/H})$)$^{a}$ & (metallicity interval)$^{b}$
    & (metallicity)$^{c}$ & (metallicity)$^{c}$ & host median$^{d}$ \\
\hline
FRB20240210A              & $0.60$ ($8.61$) & $0.34$ to $0.90$ ($8.54$ to $8.66$) & $0.51$ ($8.63$) & $0.06$ ($0.01$) & $-1.5$ \\
FRB20240312D              & $0.35$ ($8.66$) & $-0.18$ to $0.80$ ($8.56$ to $8.77$) & $0.24$ ($8.69$) & $0.06$ ($0.01$) & $-1.9$ \\
FRB20250227A$^{\dagger}$  & $1.17$ ($8.47$) & $0.56$ to $1.84$ ($8.26$ to $8.62$) & -- & $\geq 0.91$ ($\leq 8.53$) & -- \\
FRB20211127I              & $0.29$ ($8.68$) & $-0.48$ to $0.93$ ($8.53$ to $8.83$) & $-0.35$ ($8.81$) & $0.11$ ($0.02$) & $-5.6$ \\
FRB20250316A$^*$              & $1.43$ ($8.39$) & $0.64$ to $1.98$ ($8.21$ to $8.60$) & $0.97$ ($8.52$) & $0.09$ ($0.02$) & -- \\\hline
\end{tabular}
\vspace{2pt}
\begin{minipage}{\textwidth}
\footnotesize
Metallicities (in parentheses throughout) are $12+\log(\mathrm{O/H})$; O3N2
$\equiv \log(([\mathrm{O\,III}]\,\lambda5007/\mathrm{H}\beta)/([\mathrm{N\,II}]\,\lambda6584/\mathrm{H}\alpha))$.
$^{a}$Median O3N2 (metallicity) over all valid host spaxels.
$^{b}$ 95\% (2.5th--97.5th percentile) interval of the host's O3N2 (metallicity) distribution.
$^{c}$Mean and standard deviation under a Gaussian PSF mask (FWHM from Table \ref{tab:obs}) at the FRB site.
${\dagger}$: For \twotwoseven, no PSF-mask spaxel has an independent [N~\textsc{ii}] detection. We report an O3N2 lower limit (metallicity upper limit, marked $^{e}$) instead, and leave the mean/sigma-difference cells blank.
${\ast}$: For \threeonesix, the Host statistics correspond to the KCWI coverage and not the full host. We do not report the $N\sigma$ deviation as a result.
$^{d}$$(\mathrm{Mean~O3N2}_{\mathrm{FRB}} - \mathrm{Median~O3N2}_{\mathrm{host}}) /
\mathrm{Std.~dev.~O3N2}_{\mathrm{FRB}}$.
\end{minipage}
\end{table*}

\begin{figure*}
\centering
\includegraphics[width=\textwidth]{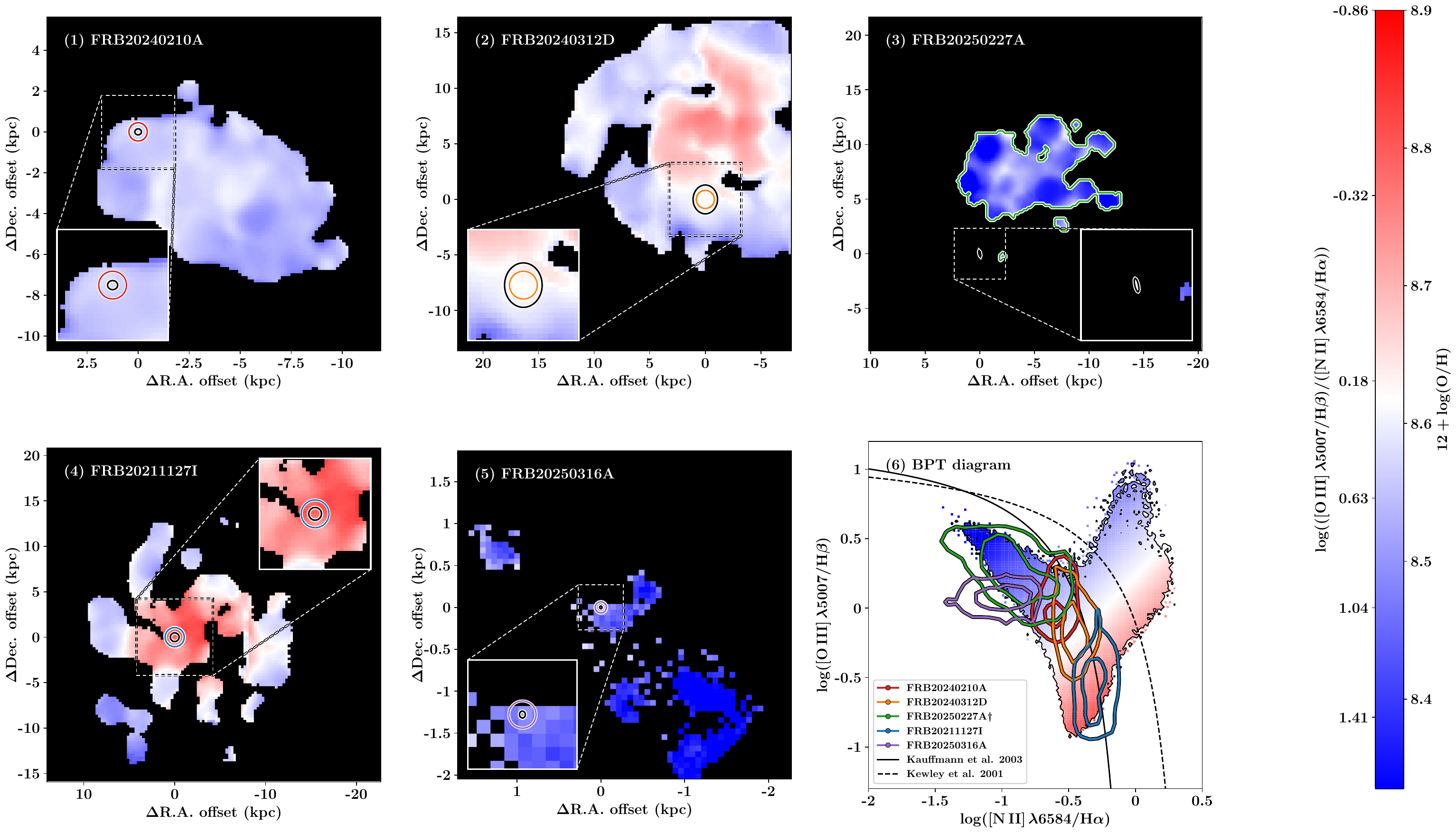}
\caption{
\textit{Panels 1-5}: Spatially resolved gas-phase metallicity maps of all four FRB host
galaxies, derived from the O3N2 index \citep{Pettini+2004} and the \citet{Nakajima+2022} calibration
(\S\ref{subsec:linefit}). The $1\sigma$ FRB localization region is shown as a black ellipse. Colored circles show the PSF FWHM centered on the FRB site. Spaxels within a region twice the PSF size, shown in the insets, inform Panel 6. In the case of \twotwoseven, since the FRB site has no \NII\ measurement, we use all the valid spaxels to inform Panel 6. \textit{Panel 6}: A BPT diagram showing 50\% and 90\% contours (with 0.1 dex smoothing along each axis) for spaxels within the colored regions from the five maps. For \twoten\, \threetwelve\, and \onetwoseven\, this is the $2\sigma$ localization region of the FRB. *: For \threeonesix\, since the localization is very precise, we use spaxels within the $6\sigma$ localization ellipse. $\dagger$: For \twotwoseven\, the FRB is significantly offset from our detected emission; hence we show contours for all spaxels with a metallicity estimate. To contextualize our measurements, we also show the same metallicity estimates in the background for SDSS galaxies. All panels share one common diverging color scale centered on the solar value
($\OH=8.69$; \citealt{Asplund+2009}) and capped at $8.9$ so that a single target's high-metallicity tail cannot compress the contrast visible in the others. The colorbar carries a dual axis: the
primary (left) axis gives the underlying O3N2 index value, and the secondary (right) axis gives the corresponding $\OH$ value.
}
\label{fig:metallicity_maps}
\end{figure*}

\begin{figure*}
\centering
\includegraphics[width=\textwidth]{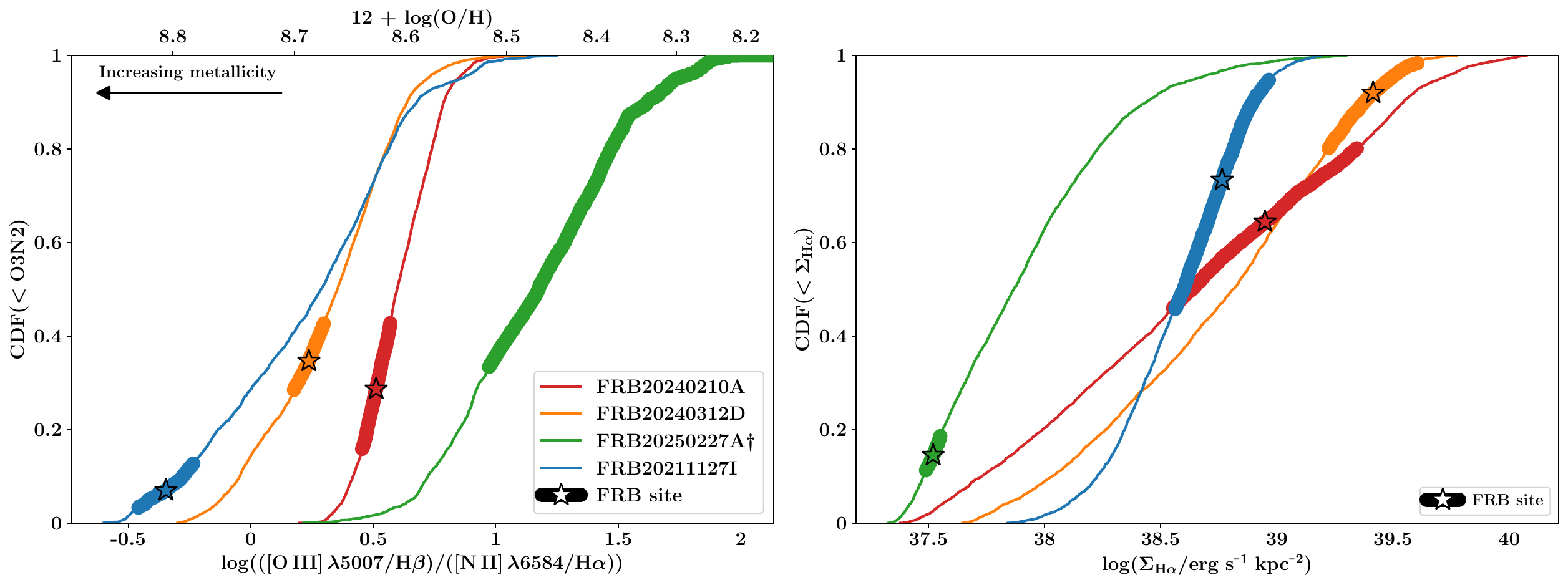}
\caption{
\textit{Left:} Cumulative distributions of the spaxel-by-spaxel O3N2 values shown in Figure~\ref{fig:metallicity_maps}, one curve per FRB host. The conversion to \OH\ is shown as a secondary horizontal axis at the top. The PSF-weighted mean value at the FRB site is shown with the black star. The thick line on each curve corresponds to the PSF-weighted standard deviation in spaxel-to-spaxel metallicity. For \twotwoseven, we are placing a $3\sigma$ lower (upper) limit on O3N2 (\OH), and hence, we show no star. We do not show the curve for \threeonesix\ as our IFU observations only cover a small portion of the galaxy. \textit{Right:} Cumulative \Hasb\ distributions with the FRB-sites in comparison, in the same way as in the left panel.} 
\label{fig:metallicity_hist}
\end{figure*}

\begin{table*}
\centering
\footnotesize
\caption{$\mathrm{H}\alpha$ surface-brightness ($\Sigma_{\mathrm{H}\alpha}$) statistics for the five FRB host galaxies.}
\label{tab:halpha_sb_stats}
\begin{tabular}{l|cc|ccc}
\hline\hline
 & \multicolumn{2}{c|}{Host} & \multicolumn{3}{c}{FRB location} \\
FRB & Median $\Sigma_{\mathrm{H}\alpha}$ & 95\% $\Sigma_{\mathrm{H}\alpha}$ interval & Mean $\Sigma_{\mathrm{H}\alpha}$ & Std. dev. $\log\Sigma_{\mathrm{H}\alpha}$ & $N\sigma$ from \\
  &  \multicolumn{2}{c|}{$(10^{38}\,\mathrm{erg\,s^{-1}\,kpc^{-2}})$} & $(10^{38}\,\mathrm{erg\,s^{-1}\,kpc^{-2}})$ & (dex) & host median \\
\hline
FRB20240210A                             & $4.32$ & $0.33$ to $65.7$ & $8.88$ & $0.40$ & $+0.8$ \\
FRB20240312D                             & $6.48$ & $0.60$ to $36.1$ & $26.0$ & $0.19$ & $+3.2$ \\
FRB20250227A$^{\dagger}$                 & $0.74$ & $0.25$ to $6.58$ & $0.33$ & $0.03$ & -- \\
FRB20211127I                             & $3.97$ & $1.11$ to $10.9$ & $5.82$ & $0.20$ & $+0.8$ \\
FRB20250316A$^{\ddagger}$                & $6.40$ & $2.97$ to $33.5$ & $7.25$ & $0.23$ & -- \\
\hline
\end{tabular}
\vspace{2pt}
\begin{minipage}{\textwidth}
\footnotesize
$\Sigma_{\mathrm{H}\alpha}$ is the $\mathrm{H}\alpha$ surface brightness, i.e., luminosity per projected kpc$^2$ at the host: $4\pi(1+z)^4$ times the observed surface brightness, converted from per steradian and per cm$^2$ to per kpc$^2$. For \threeonesix\ we use the redshift independent distance from the NASA extragalactic database (NED) \citep{Cook+2023_NED} No dust or inclination correction is applied. The data is organized between the global host statistics and FRB site statistics, similar to Table \ref{tab:metallicity_stats}.
$^{\dagger}$\twotwoseven: the corrected $\mathrm{H}\alpha$ map is fitted only where the narrowband $\mathrm{S/N}>10$, leaving 25\% of the kernel weight on spaxels without a valid detection. As a result, the scatter is likely underestimated, and thus we do not report the significance of the deviation from the host median.
$^{\ddagger}$\threeonesix: we do not report the significance of the deviation from the host median here as our footprint does not span the full host.
\end{minipage}
\end{table*}


We first assess the ionization state of gas at the FRB site to validate the use of O3N2 as a metallicity indicator. Panel 6 of Figure \ref{fig:metallicity_maps} shows the distribution of spaxels in the $2\times$ PSF FWHM centered on the FRB localization on the Baldwin, Phillips \& Terlevich (BPT) diagram \citep{BPT}. We use the full galaxy for \twotwoseven, as we are unable to measure metallicity for this FRB site. The 50\% and 90\% contours are generated from a smoothed 2D histogram of the spaxels on the diagram (smoothing scale: 0.05 dex). All five contours are generally consistent with the star-forming locus of the diagram, below the \citet{Kauffmann+2003} fit to SDSS galaxies. The centroid of \onetwoseven~is the only one of our sources lying within the ``composite'' region, i.e. above the \citet{Kauffmann+2003} curve and below the \citet{Kewley+2001} curve, indicating possible active galactic nucleus (AGN) contamination. This is further corroborated by the FRB lying close to the center of the host galaxy. That said, we find no spaxel around the FRB site lying in the AGN region of the diagram, i.e., above the \citet{Kewley+2001} curve, and thus proceed with our metallicity analysis.

The left panel of Figure \ref{fig:metallicity_hist} shows the metallicity distributions for all spaxels for which we have computed the O3N2 index. We mark the mean value within the 2$\sigma$ localization region with a vertical blue line and the standard deviation in the blue shaded region. For \twotwoseven, we place a $2\sigma$ upper limit on the metallicity at the FRB position, which is shown as a vertical dashed line with a leftward arrow. Table \ref{tab:metallicity_stats} summarizes the metallicity estimates for the host and the FRB location.

We first note observations on our methodology, i.e., spatially resolving metallicity with IFU-based strong-line diagnostics:

\begin{enumerate}
    \item The spread in \OH\ within the host, which we estimate as the 95\% interval from all spaxels, is 0.1--0.4 dex. The largest spread arises from \threeonesix\ which, despite our data only covering a small portion of the host, is still the most diverse. The host, NGC 4141, is also the closest in our sample at only 41 Mpc. Thus, we might be resolving metallicity variations better. All hosts exhibit a much greater spatial variation in metallicity than the typically reported statistical errors on \OH\ from SED fitting or longslit-based strong-line diagnostics \citep[e.g.][]{Sharma+2024, Yamasaki+2026, Dong+2026}.
    \item We measure a statistically significant deviation ($\gtrsim5\sigma$ in absolute value) from the median host value at the FRB location in the case of \onetwoseven\ (see Table \ref{tab:metallicity_stats}). \threeonesix\ also has a similar extent for the absolute deviation but only for the spatial extent of our data. For \twoten\ and \threetwelve, the relative deviation is smaller, but measured to be $1.5\sigma$ and $1.9\sigma$. Together, we conclude that spatially resolved metallicity estimates at the FRB locations provide different constraints on the FRB progenitor mechanisms than a global estimate. We elaborate on this point further in our Discussion section.
    \item The host of \twotwoseven, as is clearly seen in Figure \ref{fig:metallicity_maps}, exhibits the lowest metallicity values globally among all four hosts. Notably, \citet{Dong+2026} derived a global median value of $\OH = 8.6$ using longslit spectroscopy for the same host. This value is higher than our estimated host median value of 8.47, roughly at the 95\% interval of all spaxels with metallicity measurements in the galaxy (Table \ref{tab:metallicity_stats}). This apparent discrepancy may be driven by two main causes:
    \begin{enumerate}
        \item The longslit observation was made with a position angle of 180$^\circ$ centered on the continuum centroid of the galaxy. This geometry avoids all the low-metallicity ($\OH\lesssim8.4$) star-forming regions within the galaxy, thus biasing the metallicity estimate to higher values.
        \item The emission-line calibration adopted by \citet{Dong+2026}, from \citet{Curti+2017}, differs systematically from our choice of \citet{Nakajima+2022}. Applying our calibration choice to the same line fluxes, we obtain a value of 8.54.
    \end{enumerate}
    \item As noted in the Introduction, \citet{RBFLOAT_2025} estimate the FRB site metallicity as 8.65 for \threeonesix, but their choice of strong-line diagnostic was R23, as opposed to O3N2 in our work. This results in our estimate of 8.52 being lower by $\sim0.1$ dex, within expectations from previous studies \citep{Wisz+2025}. Interestingly, our estimate is also systematically lower than 8.62, the estimate from line fluxes reported by  \citet{Dong+2026} and applying the \citet{Nakajima+2022} calibration as we did above for \twotwoseven. This is also a $\sim5\sigma$ offset assuming the longslit data well represents the global host median.
\end{enumerate}

Next we make the following observations about the metallicity at the FRB sites:
\begin{enumerate}
    \item Intriguingly, all four hosts with FRB-site metallicity measurements in our sample range between $8.5\lesssim\OH\lesssim8.9$, while the upper limit on \twotwoseven\ places it below 8.6.
    \item The FRB-site metallicity for \onetwoseven\ is much higher than what the galaxy stellar mass might suggest if one used a median mass-metallicity relation \citep[MZR; e.g.,][]{Curti+2020}. At the host mass of $\log(M_*/M_\odot)=9.5$, the median \OH\ is 8.6 with a scatter of 0.1 dex. This places the FRB site value of 8.81 at $\sim2\times$ above the MZR scatter.
\end{enumerate}

Owing to our small sample size, we cannot claim a statistically significant preference for FRBs to occur in relatively high metallicity ($\OH\gtrsim8.5$) environments, especially since \onetwoseven\ is a clear outlier and \twotwoseven\ only has an upper limit.

The right panel of Figure \ref{fig:metallicity_hist} shows the cumulative distribution of \Hasb\ for the same four hosts. The corresponding table is Table \ref{tab:halpha_sb_stats}. Firstly, it must be noted that all five of our FRB sites coincide with \Halpha\ emitting regions. When compared against the full galaxy footprint, only \twotwoseven\ sits well below the median, while the other three lie above. Indeed, the \Hasb\ at the \threetwelve\ site is $\sim3\sigma$ above the median. While it is true that these three sites are also the top three in our sample in terms of burst-site metallicity, the two quantities are not strongly correlated. We note that we are likely underestimating the scatter in the \twotwoseven\ site \Hasb\ because nearly 25\% of the PSF kernel weight lands on spaxels with no detection. Still, this implies the burst site is at best around the 20th percentile of all \Halpha-bright spaxels.

\section{Discussion and conclusions}
\label{sec:discussion}
To summarize our study, our work utilizes data from one of the earliest set of observations with the LLAMAS IFU, and we note the instrument is very well-suited for resolving the $ < 2$ kpc environment of $z<0.05$ galaxies. The $35''\times35''$ field of view is well matched to the typical sizes of such galaxies, while the sensitivity allows us to detect \Halpha\ emission at the $10^{-18} \rm erg~s^{-1}~cm^{-2}~arcsec^{-2}$ level in $\sim1$ hour of integration. We have observed five nearby FRB hosts with IFUs and estimated the gas-phase metallicity at the burst sites.  We have compared these values against the distribution from the full host galaxy to determine if FRB sites have representative metallicities of their global host environments.

We note different metallicity estimates at FRB sites compared to global host values from our IFU observations. At least one FRB in our sample, \onetwoseven, shows significantly different values compared to the median value of the host. All four of our hosts show a spread of $\sim0.1$--$0.4$ dex in metallicity across the galaxy, which is larger than the typical statistical errors on \OH\ from SED fitting or longslit-based strong-line diagnostics. 

The metallicities we estimate at four FRB positions cover a range of values ($\OH=8.5$--8.9), and at the fifth site we place an upper limit of $\OH<8.6$. Our sample consists of non-repeaters and most recently, \citet{Dong+2026} find evidence that such FRBs may preferentially occur in hosts with {\it global} metallicity estimates of $\OH\gtrsim8.35$. We find our burst-site metallicity estimates are consistent with their work. However, the substantial metallicity variations we measure indicate that longslit spectra can produce biased estimates of metallicity depending on which regions the slit covers. This can result in a systematic error comparable to the scatter in metallicity in the hosts. Thus we urge the FRB community to take this effect into account when interpreting global metallicity distributions for FRB hosts. We strongly advocate for IFU follow-up of larger host samples to ultimately provide FRB-site relevant observational constraints on metallicity preferences for FRB occurrence.  


In addition, we find that all five of the FRB sites we examined show robust \Halpha\ detections within $\sim2$ kpc, establishing their association with recent star-formation activity. This can favor a magnetar origin for these bursts, assuming a prompt formation channel. However, within the hosts, the FRB sites do not always correspond to the most actively star-forming regions, as also echoed in studies of the UV distribution of their host light \citep{Mannings+2021, Chittidi+2021}. Translating the observed burst-site \Hasb\ to a magnetar formation rate per projected area is not a trivial exercise \citep[e.g.,][]{Beniamini+2019}. Thus we shall relegate to future work the estimation of the probability of a recent magnetar occurrence at the five burst sites.

In general, if all magnetars are equally likely to produce observable FRBs, and since most magnetars are expected to form from core-collapse supernovae (CCSNe) \citep{Xi+2025}, one may expect a wide range of burst-site values as observed by \citet{Hu+2026} ($\OH$=8.1--8.7). The onset of the CHIME/FRB Outriggers full array \citep{outriggers_2025}, DSA-2000 \citep{2019BAAS...51g.255H}, BURSTT \citep{BURSTT2022} and the FAST Core Array \citep{FASTCore2024} will enable a larger sample of nearby FRB hosts, including repeaters, in which IFU studies will be crucial to further test this. 

\section*{Acknowledgements}

S. S. gratefully acknowledges the Brinson Foundation for sponsoring his joint postdoctoral appointment at Northwestern U. and the U. of Chicago. Observations for this study were conducted via the telescope time granted by the U. of Chicago Magellan Time-Allocation Committee. We thank Prof. Robert Simcoe for for his work on engineering the LLMAS IFU and co-developing the \texttt{llamas-pyjamas} reduction package. We thank Prof. Allison Strom and Dr. Zhuyun Zhuang for the insightful discussions regarding resolved metallicity estimation. A.P.C. is a Canadian SKA Scientist and is funded by the Government of Canada / est financ\'e par le gouvernement du Canada. Y.D. is supported by the National Science Foundation (NSF) Graduate Research Fellowship under grant No. DGE-2234667. W.F. gratefully acknowledges support by the NSF under grant no. AST-2206494 and CAREER grant No. AST-2047919, the David and Lucile Packard Foundation, the Alfred P. Sloan Foundation, and the Research Corporation for Science Advancement through Cottrell Scholar Award \#28284. A.C.G. acknowledges support from The Brinson Foundation through a Brinson Prize Fellowship grant. K.W.M. is supported by NSF Grant No. 2510771. A.B.P. acknowledges support by NASA through the NASA Hubble Fellowship grant \mbox{HST-HF2-51584.001-A} awarded by the Space Telescope Science Institute, which is operated by the Association of Universities for Research in Astronomy, Inc., under NASA contract \mbox{NAS5-26555}. A.B.P. also acknowledges prior support from a Banting Fellowship, a McGill Space Institute~(MSI) Fellowship, and a Fonds de Recherche du \mbox{Qu\'ebec -- Nature} et Technologies~(FRQNT) Postdoctoral Fellowship. M.W.S. acknowledges support from the French government under the France 2030 investment plan, as part of the Initiative d'Excellence d'Aix-Marseille Universit\'e - AMIDEX (AMX-23-CEI-088). C.L. acknowledges support from the Miller Institute for Basic Research at UC Berkeley.

The Fast and Fortunate for FRB Follow-up team acknowledges support from NSF grants AST-1911140, AST-1910471, and AST-2206490. CHIME operations are funded by a grant from the NSERC Alliance Program and by support from McGill University, University of British Columbia, and University of Toronto. CHIME/FRB Outriggers are funded by a grant from the Gordon \& Betty Moore Foundation. We are grateful to Robert Kirshner for early support and encouragement of the CHIME/FRB Outriggers Project, and to Dusan Pejakovic of the Moore Foundation for continued support. The CHIME/FRB Project was funded by a grant from the CFI 2015 Innovation Fund (Project 33213) and by contributions from the provinces of British Columbia and Québec, and by the Dunlap Institute for Astronomy and Astrophysics at the University of Toronto. Additional support was provided by the Canadian Institute for Advanced Research (CIFAR), the Trottier Space Institute at McGill University, and the University of British Columbia. 

\software{Numpy, Scipy, Astropy, Matplotlib, LLAMAS-Pyjamas, PypeIt. In addition, Claude Code was used to significantly augment all aspects of this work; from literature review to analysis to drafting the manuscript.}

\appendix
\section{Post-processing of reduced science exposures}
\label{appendix:postprocessing}
After basic extraction and wavelength calibration of the individual fiber spectra into row-stacked-spectra (RSS) format for each of LLAMAS's three spectral channels (blue, green, and red), each science exposure was converted to a datacube using the instrument's fiber-to-sky mapping. This section describes the three post-processing steps applied to those cubes before scientific analysis: absolute flux calibration (\S\ref{subsec:fluxcal}), multi-exposure spatial registration and combination (\S\ref{subsec:combine}), and astrometric calibration onto an archival imaging reference frame (\S\ref{subsec:astrometry}).

\subsection{Flux Calibration}
\label{subsec:fluxcal}

Absolute flux calibration was derived from observations of the CALSPEC spectrophotometric standard star HD~14943 \citep{Bohlin2014, Bohlin2020}, obtained on the same night as the science exposures. The standard's spectrum was extracted from the RSS data using a circular aperture centered on the fiber-weighted centroid of its continuum light; the aperture radius was chosen large enough that a curve-of-growth analysis of the star's white-light image showed it recovered all ($>98\%$) of the total flux, so aperture losses do not contribute meaningfully to the calibration uncertainty.

The extracted count spectrum was compared against the star's tabulated CALSPEC flux standard to derive a sensitivity function independently for each of the three channels, following standard long-slit/fiber spectrophotometric calibration practice \citep{Oke1990, Hamuy1992} as implemented in the open-source spectroscopic reduction package \textsc{PypeIt} \citep{pypeit}. Each exposure's airmass and integration time were propagated into the calibration together with an atmospheric extinction curve appropriate for the observing site,
placing the resulting sensitivity functions on an absolute physical scale (erg~s$^{-1}$~cm$^{-2}$~\AA$^{-1}$) independent of the specific conditions of any individual exposure. The appropriate per-channel sensitivity function was then applied to every spaxel of each science exposure's datacube prior to spatial combination (\S\ref{subsec:combine}).

\subsection{Cube Alignment and Combination}
\label{subsec:combine}

Each target was observed as a sequence of several individual exposures, splitting the total integration to reject cosmic rays. Before combining, we spatially registered the per-exposure cubes onto a common grid. For each target, we solved this registration using the green-channel exposures by cross-correlating each exposure's broadband (white-light) continuum image against that of a designated reference exposure to measure its spatial offset. The resulting per-exposure astrometric corrections were then applied uniformly to the corresponding blue- and red-channel cubes from the same exposure, ensuring that all three channels of a given exposure share one internally consistent spatial registration, even though the alignment itself is solved only once, on the channel best suited to it.

Once registered onto the common grid, the flux-calibrated exposures for each channel were combined via inverse-variance weighting using the pipeline's propagated per-spaxel variance, producing a single cosmic-ray-rejected, depth-optimized datacube per spectral channel for each target. The reference exposure's astrometric grid and field footprint were adopted as those of the final combined cube.

\subsection{Astrometric Calibration and Registration to Archival Imaging}
\label{subsec:astrometry}

Since the pipeline-derived WCS solution was inaccurate, the combined cube at this stage was not correctly aligned to existing imaging. We took the following steps to rectify this:
\begin{enumerate}
    \item We visually examined each combined cube and reoriented it by flipping the horizontal axis and subsequently rotating clockwise or anti-clockwise (as required) by 90$^\circ$ into a standard north-up, east-left orientation.
    \item With further visual examination of the continuum image shown in Figure \ref{fig:narrowband} we adjusted the \texttt{CRVAL} keywords to ensure the continuum centroid roughly matched the galaxy position in the catalog of the Pan-STARRS or Legacy Survey reference image.
    \item We further refined the offset by subsequently cross-correlating the \Halpha\ emission-line map (the \Hbeta\ map for FRB\,20211127I, whose \Halpha\ map shows banding) against the same reference image within a search window of radius 1.5 times the FWHM of the galaxy's continuum emission, centered on the coarse solution. The line map is clumpier, allowing us to identify substructure in emission that could also be seen in the legacy imaging. The cross-correlation peak is located on the reference image's pixel grid, so the residual astrometric uncertainty is of order half a reference pixel ($\approx$0.15'', or $\approx$50\% of a spaxel).
\end{enumerate}

This procedure ties each combined cube's astrometric solution directly to the Gaia reference frame \citep{Lindegren+2018, GaiaDR2, GaiaDR3} through the Pan-STARRS or Legacy Survey reference images, providing the relative astrometric precision needed to localize FRBs with respect to resolved host-galaxy substructure such as individual star-forming regions. We show the quality of our alignment in Figure \ref{fig:sample} where the \Halpha\ contour neatly align with the galaxy substructure apparent in the broadband images.



\bibliography{reference}{}
\bibliographystyle{aasjournalv7}



\end{document}